\documentclass[runningheads]{llncs}
\usepackage[pdfstartview=XYZ,
bookmarks=true,
colorlinks=true,
linkcolor=blue,
urlcolor=blue,
citecolor=blue,
pdftex,
bookmarks=true,
linktocpage=true, 
hyperindex=true
]{hyperref}

\usepackage{orcidlink}

\usepackage{comment}
\usepackage{soul}
\usepackage{xcolor}
\usepackage{subfigure}
\usepackage{amssymb}
\usepackage{amsmath}
\usepackage{amsfonts}
\usepackage{graphicx}
\usepackage{bm}
\usepackage{xfrac}
\usepackage{hyperref}
\usepackage[acronym,shortcuts, hyperfirst=false]{glossaries}
\usepackage{balance}
\usepackage{caption}
\usepackage{bbm}
\usepackage{paralist}
\usepackage{subfigure}
\usepackage{multicol}

\newacronym{hywban}{HyWBAN}{Hybrid Wireless Body Area Network}
\newacronym{snr}{SNR}{Signal-to-Noise Ratio}
\newacronym{BLE}{BLE}{Bluetooth Low Energy}
\newacronym{vlc}{VLC}{Visible Light Communication}
\newacronym{nlos}{NLOS}{Non-Line-Of-Sight}
\newacronym{los}{LOS}{Line-Of-Sight}
\newacronym{rf}{RF}{Radio Frequency}
\newacronym{led}{LED}{Light Emitting Diodes}
\newacronym{owc}{OWC}{Optical Wireless Communications}
\newacronym{wban}{WBAN}{Wireless Body Area Network}
\newacronym{wce}{WCE}{Wireless Capsule Endoscope}
\newacronym{uwb}{UWB}{ultra wideband}
\newacronym{nb}{NB}{narrowband}
\newacronym{ISM}{ISM}{Industrial, Scientific and Medical}
\newacronym{hbc}{HBC}{Human Body Communications}
\newacronym{csi}{CSI}{Channel State Information}
\newacronym{iot}{IoT}{Internet of Things}
\newacronym{ir}{IR}{Infrared}
\newacronym{nir}{NIR}{Near-Infrared}
\newacronym{pd}{PD}{Photo-Detector}
\newacronym{dl}{DL}{Deep Learning}
\newacronym{mitm}{MitM}{Man-in-the-Middle}
\newacronym{sc}{SC}{Semantic Communications}
\newacronym{psd}{PSD}{power spectral density}
\newacronym{udp}{UDP}{User Datagram Protocol}
\newacronym{pls}{PLS}{Physical Layer Security}

\usepackage{cuted}
\usepackage{upgreek}
\usepackage{booktabs}

\usepackage{tabularx}
\usepackage{threeparttable} 
\usepackage{multirow} 
\usepackage{rotating} 
\usepackage{bigstrut} 
\usepackage{setspace}
\usepackage{tablefootnote}
 \usepackage{tabularray}

\newcommand{\parag}[1]{\noindent\textbf{#1. }}

\usepackage{algpseudocode}
\usepackage[ruled,vlined,noend]{algorithm2e}
\SetKwInput{kwFailure}{Failure}

\usetikzlibrary{svg.path}
\definecolor{orcidlogocol}{HTML}{A6CE39}
\tikzset{
	orcidlogo/.pic={
		\fill[orcidlogocol] svg{M256,128c0,70.7-57.3,128-128,128C57.3,256,0,198.7,0,128C0,57.3,57.3,0,128,0C198.7,0,256,57.3,256,128z};
		\fill[white] svg{M86.3,186.2H70.9V79.1h15.4v48.4V186.2z}
		svg{M108.9,79.1h41.6c39.6,0,57,28.3,57,53.6c0,27.5-21.5,53.6-56.8,53.6h-41.8V79.1z M124.3,172.4h24.5c34.9,0,42.9-26.5,42.9-39.7c0-21.5-13.7-39.7-43.7-39.7h-23.7V172.4z}
		svg{M88.7,56.8c0,5.5-4.5,10.1-10.1,10.1c-5.6,0-10.1-4.6-10.1-10.1c0-5.6,4.5-10.1,10.1-10.1C84.2,46.7,88.7,51.3,88.7,56.8z};
	}
}
\newcommand\orcidicon[1]{\href{https://orcid.org/#1}{\mbox{\scalerel*{
				\begin{tikzpicture}[yscale=-1,transform shape]
					\pic{orcidlogo};
				\end{tikzpicture}
			}{|}}}}

\newcommand{\simone}[1]{\textcolor{black}{#1}}

\newcommand{\matti}[1]{\textcolor{cyan}{#1}}

\usepackage{todonotes}
\newboolean{showcomments}
\setboolean{showcomments}{true}         

\ifthenelse{\boolean{showcomments}}
{\newcommand{\nb}[2]{
		\fbox{\bfseries\sffamily\scriptsize#1}
		{\sf\small$\blacktriangleright$\textit{\textcolor{red}{#2}}$\blacktriangleleft$}
	}
}
{\newcommand{\nb}[2]{}
	
}

\newcommand\ssnote[1]{\nb{SS}{#1}}

\newcommand\mhnote[1]{\nb{MH}{#1}}

\begin{document}
%
\title{{Securing Hybrid Wireless Body Area Networks (HyWBAN): Advancements in Semantic Communications and Jamming Techniques}}
\titlerunning{Securing Hybrid Wireless Body Area Networks (HyWBAN)}
%
\author{Simone Soderi\inst{1}\orcidlink{0000-0002-1024-9470}, 
Mariella Särestöniemi\inst{2,3}\orcidlink{0000-0003-4691-7570},
Syifaul Fuada\inst{3}\orcidlink{0000-0002-5258-5149},\\
Matti Hämäläinen\inst{3}\orcidlink{0000-0002-6115-5255},
Marcos Katz\inst{3}\orcidlink{0000-0001-9901-5023},
Jari Iinatti\inst{3}\orcidlink{0000-0001-6420-1547}
}
%
\authorrunning{S. Soderi \textit{et al.}}

%
\institute{IMT School for Advanced Studies Lucca, Lucca, Italy\\
\email{\texttt{simone.soderi@imtlucca.it}\\} \and
 Health Sciences and Technology, Faculty of Medicine, University of Oulu, \\ Oulu, Finland \\
 \email{\texttt{mariella.sarestoniemi@oulu.fi}\\}
 \and
Centre for Wireless Communications,  Faculty of Information Technology and Electrical Engineering, University of Oulu, Oulu, Finland \\
\email{\texttt{\{syifaul.fuada,matti.hamalainen,marcos.katz,jari.iinatti\}@oulu.fi}\\}
}
%
\maketitle              
\begin{abstract}
This paper explores novel strategies to strengthen the security of Hybrid Wireless Body Area Networks (HyWBANs), essential in smart healthcare and Internet of Things (IoT) applications. Recognizing the vulnerability of HyWBAN to sophisticated cyber-attacks, we propose an innovative combination of semantic communications and jamming receivers. This dual-layered security mechanism protects against unauthorized access and data breaches, particularly in scenarios involving in-body to on-body communication channels. We conduct comprehensive laboratory measurements to understand hybrid (radio and optical) communication propagation through biological tissues and utilize these insights to refine a dataset for training a Deep Learning (DL) model. These models, in turn, generate semantic concepts linked to cryptographic keys for enhanced data confidentiality and integrity using a jamming receiver. The proposed model demonstrates a significant reduction in energy consumption compared to traditional cryptographic methods, like Elliptic Curve Diffie-Hellman (ECDH), especially when supplemented with jamming. Our approach addresses the primary security concerns and sets the baseline for future secure biomedical communication systems advancements.

\keywords{Heterogeneous  \and WBAN \and energy \and security \and optical \and RF \and near-infrared communications.}
\end{abstract}

\newpage
\noindent\rule{8.4cm}{1pt}\\
Please cite this version of the paper:\\

S. Soderi, M. Särestöniemi, S. Fuada, M. Hämäläinen, M. Katz, and J. Iinatti. “Securing Hybrid Wireless Body Area Networks (HyWBAN): Advancements in Semantic Communications and Jamming Techniques”. In: \textit{2024 Nordic Conference on Digital Health and Wireless Solutions, NCDHWS 2024}, May 2024, doi:10.1007/978-3-031-59091-7\_24
\\
You may use the following bibtex entry:
\begin{verbatim}
@inproceedings{Soderi2024SecuringHyWBAN,
  title={Securing Hybrid Wireless Body Area Networks (HyWBAN): 
  Advancements in Semantic Communications and Jamming Techniques},
  author={Soderi, Simone and Särestöniemi, Mariella and Fuada, 
  Syifaul and Hämäläinen, Matti and  Katz, Marcos and Iinatti, Jari},
  booktitle={Digital Health and Wireless Solutions},
  pages={1--19},
  year={2024},
  organization={Springer},
  doi={10.1007/978-3-031-59091-7_24}
}
\end{verbatim}
\noindent\rule{8.4cm}{1pt}
%
%
\section{Introduction} \label{SEC:INTRO}



\simone{The advent of wireless and mobile communications technologies has been essential in enhancing healthcare, marking a paradigm shift towards more proactive and personalized medical interventions. The concept of smart healthcare is at the forefront of this transformation, offering many opportunities to address the growing needs of an ageing population and the increasing prevalence of chronic diseases~\cite{2023:smart-health_6G}.
Remote health monitoring, a cornerstone of modern healthcare, has emerged as a cost-effective and efficient approach to disease prevention and healthcare provision, especially with the integration of 5G and 6G technologies. These advancements are pivotal in supporting in-body communications with implanted medical devices, enabling real-time health provisioning, virtual consultations, better diagnostics, and telesurgeries, among other benefits~\cite{2023:smart-health_6G}.
Historically, information transmission through biological tissues has predominantly relied on radio and acoustic waves~\cite{2020:inbody-light-marcos}. However, these conventional methods are fraught with challenges, including security, safety, privacy, and interference, necessitating the exploration of alternative communications media. The vulnerabilities of implantable or in-body devices to hacks and unauthorized access have underscored the urgent need for enhanced security measures~\cite{2021:bio-cyber-security,2017:cybersec_med_guide}.}

\ac{owc} has emerged as a promising alternative, utilizing light, especially in the near-infrared range, to transmit information through biological tissues. This method offers many advantages, including high security, privacy, safety and low complexity, as well as low power consumption. It has been used to successfully establish connectivity to electronic devices embedded under the skin~\cite{2020:connectivity-inbody-uwb}. Going further, a hybrid solution which is merging both radio-based and optical-based technologies in \ac{wban} context can open a new, more secure way to implement personalized healthcare services and transfer personal health data. This is also a way to reduce radio signal emission towards the human body. 

\simone{\acp{hywban} stand at the forefront of innovation in healthcare and \ac{iot} applications, merging radio and \ac{owc}. These networks offer remarkable advantages such as data throughput enhancement, enhanced security, and improved reliability, making them ideal for critical healthcare applications and various services, from patient monitoring to advanced diagnostics.}
\simone{Reconfigurable \acp{hywban} takes adaptability to the next level with a dynamic architecture, ensuring consistent performance in diverse environments. Preliminary studies on \acp{hywban} underscore their potential, showcasing notable performance and energy efficiency improvements~\cite{2019:reconfig_optical_radio}.}
\simone{Energy harvesting is crucial to \acp{hywban}~\cite{2022:analysis-hybird-visible-IR}, focusing on developing energy-autonomous nodes that enhance sustainability and reduce maintenance. The networks' advanced sensing capabilities support single and dual-mode sensing, enabling comprehensive data collection for diverse applications. Moreover, \acp{hywban}' design promotes sustainable operations, which is essential in today's environmentally conscious landscape.}
\simone{Optimized data transmission functionality in \acp{hywban} caters to the high demands of medical and \ac{iot} applications. A significant feature is the ability to transfer energy to in-body devices, ensuring continuous operation. Additionally, \acp{hywban}' advanced sensing capabilities are essential in medical diagnostics, allowing for detailed tissue analysis and health monitoring, thus revolutionizing healthcare and \ac{iot} applications~\cite{2023:smart-health_6G}.}

Over the years, academia has shown interest in \ac{pls} solutions that aim to protect communications by exploiting the properties of the communication media~\cite{bloch2011physical,2017:soderi_wbplsec,2014:soderi-fingerprint,2022:soderi_pls_vlc_rgb}. These techniques consist of processing the signal sent over a channel in such a way as to obtain certain security properties without resorting to specific primitives, typically cryptography, offered by layers above the physical level.
In this paper, we show how to combine \ac{pls} techniques with DL algorithms to improve the security of \acp{hywban}.

%
%
\begin{figure}[!t]
	\centering
	\includegraphics[width=0.8\columnwidth]{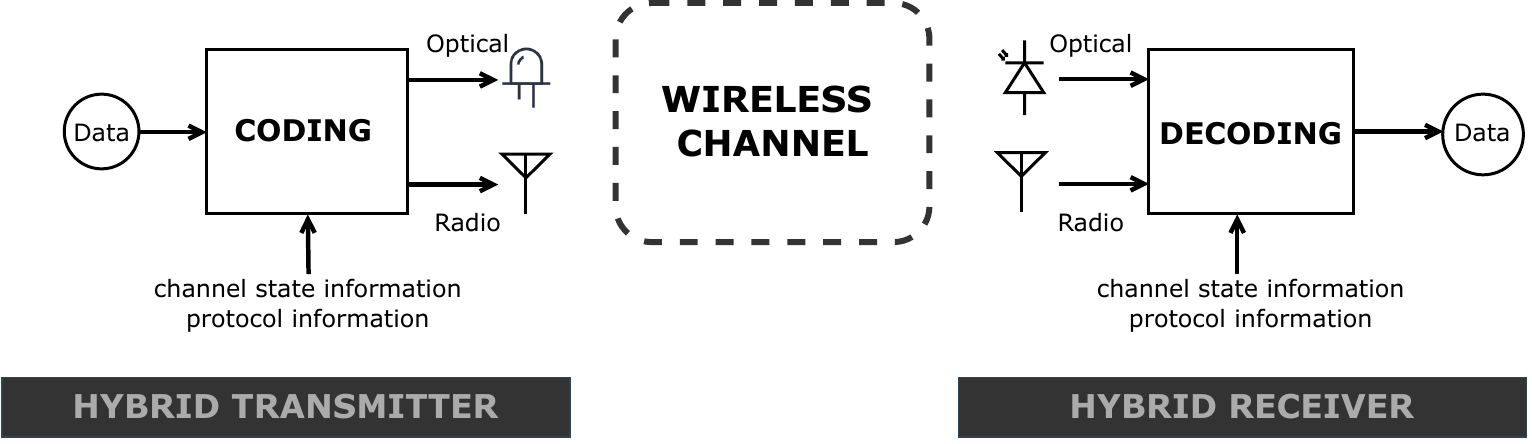}
	\caption{Coding strategy for hybrid networks.}
	\label{fig:hybrid-classic}
\end{figure}
\parag{Motivation}\simone{In this article, we have decided to use a model already used in the literature that involves defining operating modes based on the combinations of the two communications channels~\cite{2019:reconfig_optical_radio,2020:vlc-radio-opportunities}. Fig.~\ref{fig:hybrid-classic} depicts how hybrid radio-optical wireless networks utilize Shannon's theory, which defines the maximum channel capacity for communications. It shows the dynamic selection of the device's operating modes based on factors like channel state information (radio/optical) and user context.}
\simone{In our view, \acp{hywban} can improve \textit{security} by integrating in the radio transmissions the \ac{owc}, which is known for the localised and secure transmission of signals. Encoding signals across radio and optical channels maximises secrecy, in line with recent theoretical work on conventional networks. This approach exploits the inherent security features of optical communications, addressing the vulnerabilities of \acp{wban}. This paper uses data measured in the laboratory to implement an \textit{innovative hybrid network security scheme} using semantic communications and intentional interference.}

\parag{Contribution} \simone{In particular, hybrid communications have been of great interest for sensor networks. In a digital healthcare scenario, protecting these communications and doing so effectively while consuming as little energy as possible makes significance. The contributions of this article can be summarised as follows. \textit{(i)} We present a novel concept that exploits the combination of \ac{sc} with a jamming receiver to improve the confidentiality and integrity of these wireless communications. \textit{(ii)} We performed measurements in the laboratory to study the propagation of hybrid (radio and optical) communications in biological tissue. These measurements allowed us to define part of the dataset used for the \ac{dl}  model. Finally, \textit{(iii)} we evaluated and performed a security analysis of the \acp{hywban}.}

\simone{The remainder of the paper is organized as follows. Section~\ref{SEC:Background} briefly recalls the concepts useful for understanding the paper. 
Section~\ref{sec:security_analysis} discusses the major security threats in this scenario. Then, Section~\ref{SEC:enhancing-security} presents the proposed scheme to enhance the security of hybrid networks. Section~\ref{sec:results} presents the results achieved in terms of performance and energy cost. 
Finally, Section~\ref{SEC:Conclusions} concludes the paper by discussing our findings.}

%
%
\section{Background} \label{SEC:Background}
\simone{This section introduces the radio and optical technologies used for wireless sensor communication. The aim is to provide some notions before discussing how hybrid networks combine these two technologies.}
\subsection{Radio-based WBAN technologies} \label{SEC:radio-based-background}
\begin{figure}[!b]
	\centering
	\includegraphics[width=0.6\columnwidth]{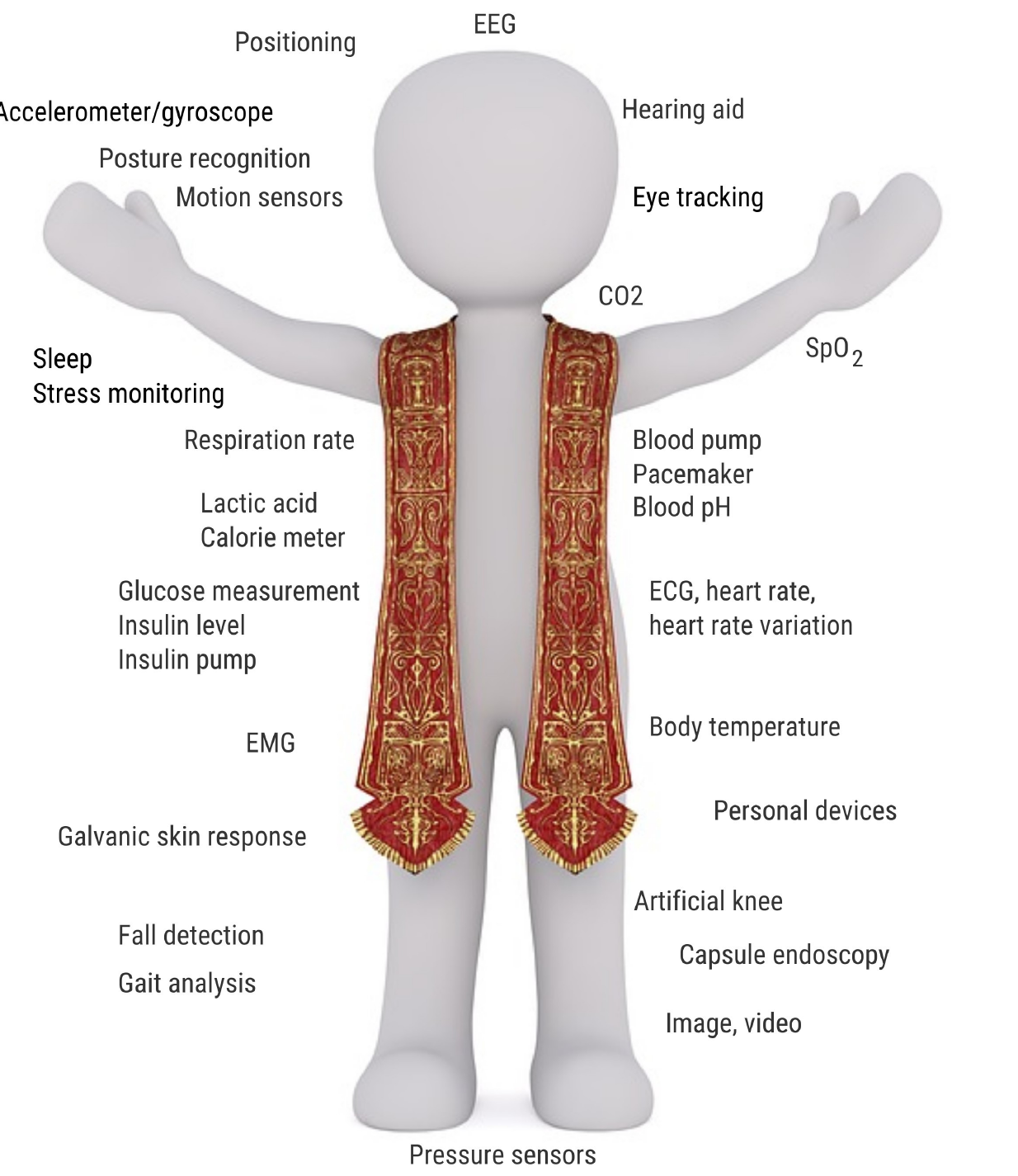}
	\caption{Variability of the possible sensors that can be used in the \ac{wban} context.}
	\label{fig:wban_ukkeli}
\end{figure}

\ac{wban} is a way to link various wearable sensor nodes wirelessly into one individual and personalized network used to monitor a person's psycho-physiological vital signs. Depending on the need, the vital sensors can be distributed all around the human body. Low-power consumption, small size, and lightweight are the requirements set for the nodes to enable user acceptance. In principle, the amount of connected sensors within one \ac{wban} can be high, but a realistic number is less than five for the sake of usability. The basic idea behind a \ac{wban} is that dedicated sensors are collecting vital information and transmitting it wirelessly to the central node (called a hub), which then pre-process the data or conveys it further. Fig.~\ref{fig:wban_ukkeli} shows the variability of the vital sensor nodes, which can be used in the \ac{wban} context (the list of sensors is not exhaustive)~\cite{2023:ismict}. In addition to sensors which are attached to the skin, so called on-body sensors, \ac{wban} can utilize smart implants, such as pacemakers, or other in-body sensors/devices, such as \ac{wce}. In \ac{wban}, all the nodes are connected to the on-body hub to enable real-time information transmission towards backbone infrastructure. Typically, \ac{wban} is using a one-hop star network topology. 

%
%

Currently, \textit{de facto} wireless standard in \ac{wban} is \ac{BLE} but there are also other dedicated \ac{wban} standards available, such as ETSI SmartBAN~\cite{2020:smartban}, IEEE 802.15.6~\cite{2012:802156}, or IEEE 802.15.4~\cite{2015:802154}. The latter one is better known via its higher layer protocols ZigBee and 6LoWPAN. 

From a radio technology point-of-view, \ac{wban} connectivity can be based on \ac{nb} signals, which are used, e.g., in \ac{BLE} and SmartBAN, or \ac{uwb}, which is adopted by~\cite{2012:802156}. The most common frequency band at the moment for \ac{nb} signal is \ac{ISM} band at about 2.4 GHz. On the other hand, e.g.,~\cite{2012:802156} defines several \ac{nb} frequency bands for \ac{wban} use also occupying sub-GHz frequencies. As operating in a highly populated frequency range, \ac{ISM} band is typically subject to high interference originated from other radio equipment nearby. 
The selected frequency band, as well as signal bandwidth, also have an impact on the observed signal propagation properties through/along tissues, positioning accuracy, throughput, etc., depending on the application and use-case. If high resolution, real-time data is needed, then \ac{uwb} can be the best option from radio-based technologies. Lower performance requirements and deeper in-body penetration, however, are favouring \ac{nb} technology. Reference~\cite{2012:802156} also defines \ac{hbc} technology operating around 21 MHz, but this is omitted in this review due to its deviation from the conventional \ac{rf}-based communications as being a coupling-based solution. 

The original network topology in \ac{wban} is based on a star topology, where all the data flows are going through the central node, the hub. In this case, the hub is also a bridge from the body domain to the backbone network. The recent development, e.g., in ETSI SmartBAN has introduced and defined a hub-to-hub communications to transfer information between adjacent \ac{wban} networks~\cite{2023:H2H}. In addition, a two-hop relay functionality is included in the SmartBAN technical specification~\cite{2023:relay}. From the security viewpoint, all the hops between \ac{wban} nodes, although being short, should be reliable but also secure as the communications chain is as reliable as its weakest link. This highlights the importance of light security protocols to be used also in the \ac{wban} context.

%
%
\subsection{Optical communications in Wireless Sensor Networks} \label{SEC:optical-comm-background}
\simone{The utilization of optical communications, including \ac{vlc} and \ac{ir} technologies, in wireless sensor networks is gaining interest for various \ac{iot} and body network applications. Optical communication offers security, bandwidth, and energy efficiency advantages, which are crucial for IoT deployments.}

\simone{\ac{vlc} utilizes \ac{led} to transmit data using the visible spectrum. This approach is inherently secure due to the limited light propagation and offers high data rates, making it suitable for indoor \ac{iot} applications \cite{2021:Light-IoT}. \ac{vlc}'s potential in hybrid optical-wireless networks for next-generation communications, especially in 5G and beyond, is highlighted in \cite{2018:optical-hybird-5G}.
\ac{ir} communication leverages the non-visible spectrum for data transmission, offering benefits in terms of device miniaturization and reduced interference with existing \ac{rf} systems. Its suitability for low-data-rate \ac{iot} applications, especially in hybrid networks combining \ac{ir} and \ac{vlc}, is explored in \cite{2022:analysis-hybird-visible-IR}. \ac{ir} in hybrid radio-optical wireless networks offers innovative solutions for versatile \ac{iot} applications, as discussed in \cite{2023:owc-IoT-datarate}.
Integrating optical and wireless technologies in a hybrid framework opens new avenues for enhancing \ac{iot} network performance. The synergy of \ac{rf} and optical communication technologies in hybrid networks is investigated in \cite{2017:heterogenous-radio-optical}, which outlines the implementation and advantages of such an approach.}

\section{Security Analysis of HyWBANs} \label{sec:security_analysis}
\simone{Developing next-generation networks to support better biomedical applications presents an opportunity. However, cyber-security risks arise mainly from this technology's highly interconnected and ubiquitous nature~\cite{2023:smart-health_6G}. 
Therefore, the cybersecurity analysis of these hybrid communications begins with choosing a system model that best represents the problem.}
%
%
\begin{figure}[!t]
	\centering
	\includegraphics[width=0.75\columnwidth]{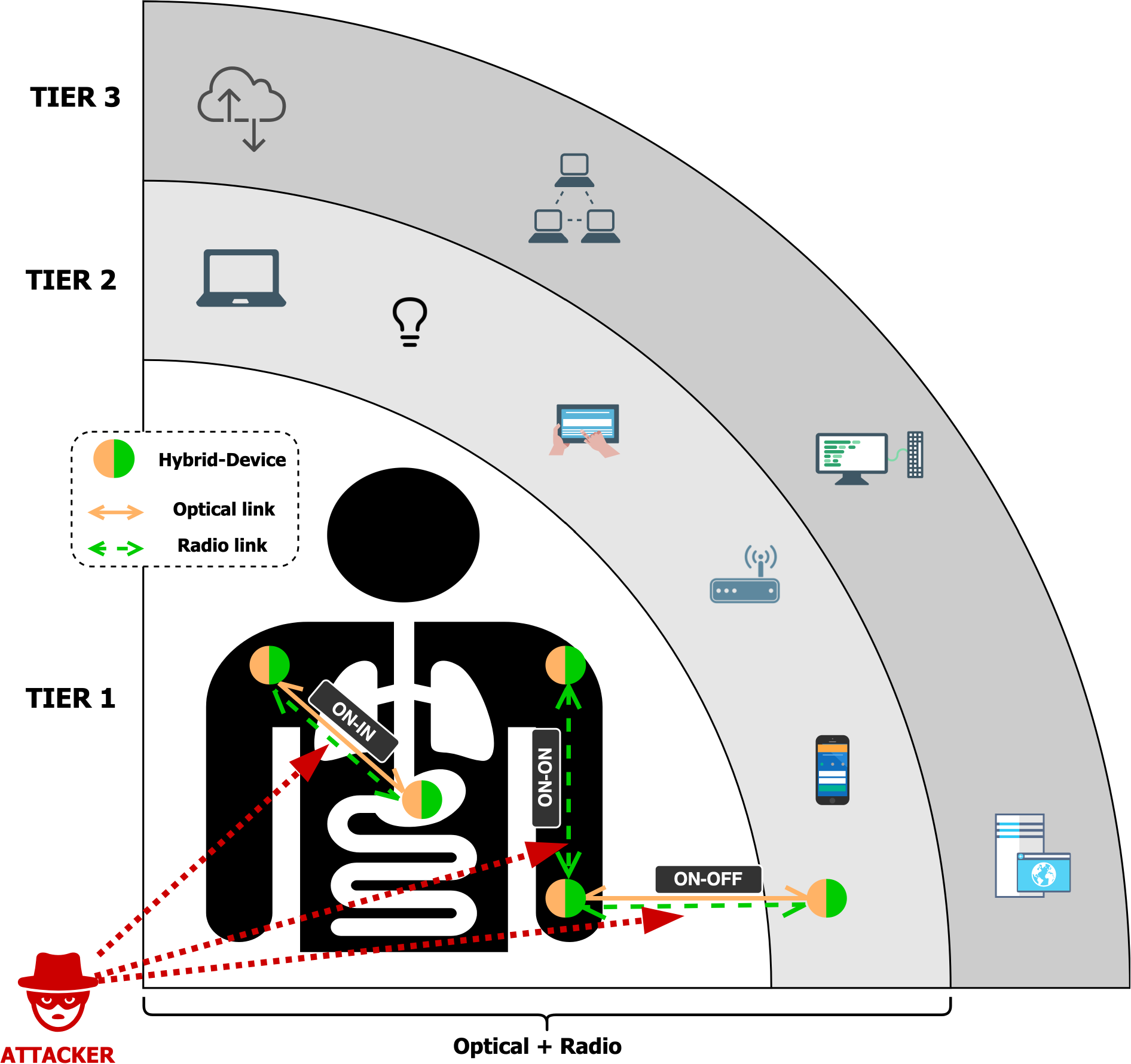}
	\caption{Communications tiers system model, in which hybrid communications operate in the first two tiers.}
	\label{fig:tier-model}
\end{figure}

\simone{\acp{wban} are components of cyberspace that assist people in their daily activities and collect data from persons. \acp{wban} and, more broadly, wearable wireless networks (WWNs) have three communication layers, according to the \textit{tier model}~\cite{2005:arch-health-monitor}.
As shown in  Fig.~\ref{fig:tier-model}, wearable sensors capture data in Tier~1 and transmit it to Tier~2 for aggregation and data processing. Finally, data is sent to Tier~3 and made available for remote access. 
The \acp{hywban} follow the same system model, where radio link, optical link, or both can be used for each type of communications in Tier~1 and Tier~2.
As illustrated in Fig.~\ref{fig:tier-model}, we have different types of communications in \ac{hywban}: on-body to in-body devices (labelled as \textit{On-In})  and on-body to on-body (labelled as \textit{On-On}) devices that operate in Tier~1. Instead, at Tier~2, all communications are off-body, including on-body to off-body devices (labelled \textit{On-Off}).
We assume that \acp{hywban} operate up to Tier~2, as depicted in Fig.~\ref{fig:tier-model}.}

\simone{One of the main security problems of this communication chain is that Eve, the adversary (attacker) shown in Fig.~\ref{fig:tier-model}, can carry out several attacks. We can assume that she has complete control to intercept and modify all messages exchanged between \ac{hywban} nodes~\cite{1983:dolev-yao_model}. In the rest of the paper, we analyse the possible attacks and their mitigation.}

%
%
\subsection{Security Threats Overview}  \label{SUBSEC:threat-models}
The complexity of \acp{hywban}, which combines \ac{rf} and \ac{owc}, far exceeds traditional communications systems due to their dual-channel nature. This sophistication poses significant challenges for attackers attempting to compromise the network, as they must navigate radio and optical channels.
%
%
\begin{figure}[!b]
	\centering
	\includegraphics[width=0.75\columnwidth]{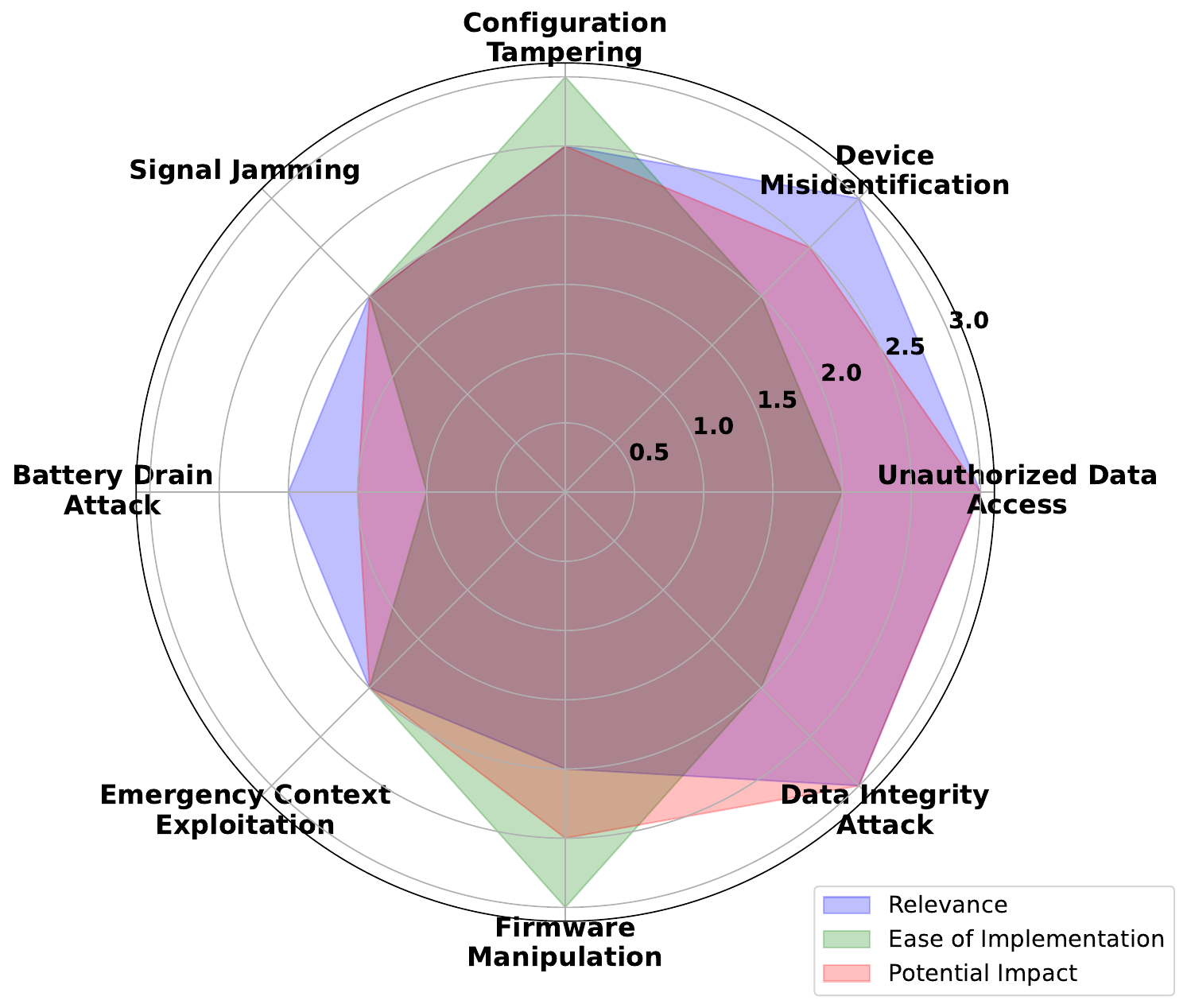}
	\caption{Security threats to \ac{hywban}.}
	\label{fig:radar-chart}
\end{figure}

In the HyWBANs domain, the security of communication channels is essential, especially when considering transmitting sensitive health data. 
This section delves into the nuanced vulnerabilities inherent to \ac{rf} and \ac{owc}, laying the groundwork for understanding the superior security posture of optical communications in specific scenarios. 
RF communications, by their very nature, are susceptible to eavesdropping due to their omnidirectional signal propagation. This characteristic allows malicious entities to intercept signals without necessitating a direct line of sight, thereby posing a significant risk to the confidentiality of transmitted data. Conversely, optical communications demand a line-of-sight for effective transmission, inherently restricting the potential for unauthorized interception. Despite this advantage, optical channels are not impervious to security threats. A breach in the line-of-sight or sophisticated techniques to capture reflected optical signals can compromise data integrity and confidentiality. 

In the context of \acp{hywban}, an attacker would primarily focus on tactics that enable eavesdropping on biomedical device communications. These tactics could include exploiting network security protocol vulnerabilities, conducting \ac{mitm} attacks to intercept data, or using sophisticated techniques to bypass encryption. Reconnaissance plays a crucial role, as the attacker must gather detailed information about the network's configuration and security mechanisms to successfully deploy malware or other attack vectors.

\simone{Fig.~\ref{fig:radar-chart} presents a multifaceted evaluation of security threats in \acp{hywban}. Each threat is analyzed based on three critical dimensions: \textit{relevance} to the network's security, \textit{ease of implementation} by potential attackers, and the \textit{potential impact} on network integrity and functionality. This brief assessment enables a slight understanding of each threat's effectiveness and helps prioritize security measures.}

To fortify the security framework of HyWBANs against these vulnerabilities, we introduce a semantic communication method that significantly enhances the security of transmitted data. 

\section{Enhancing the Security of HyWBAN through Semantic Communications}  \label{SEC:enhancing-security}
\begin{figure}[!b]
	\centering
	\includegraphics[width=0.95\columnwidth]{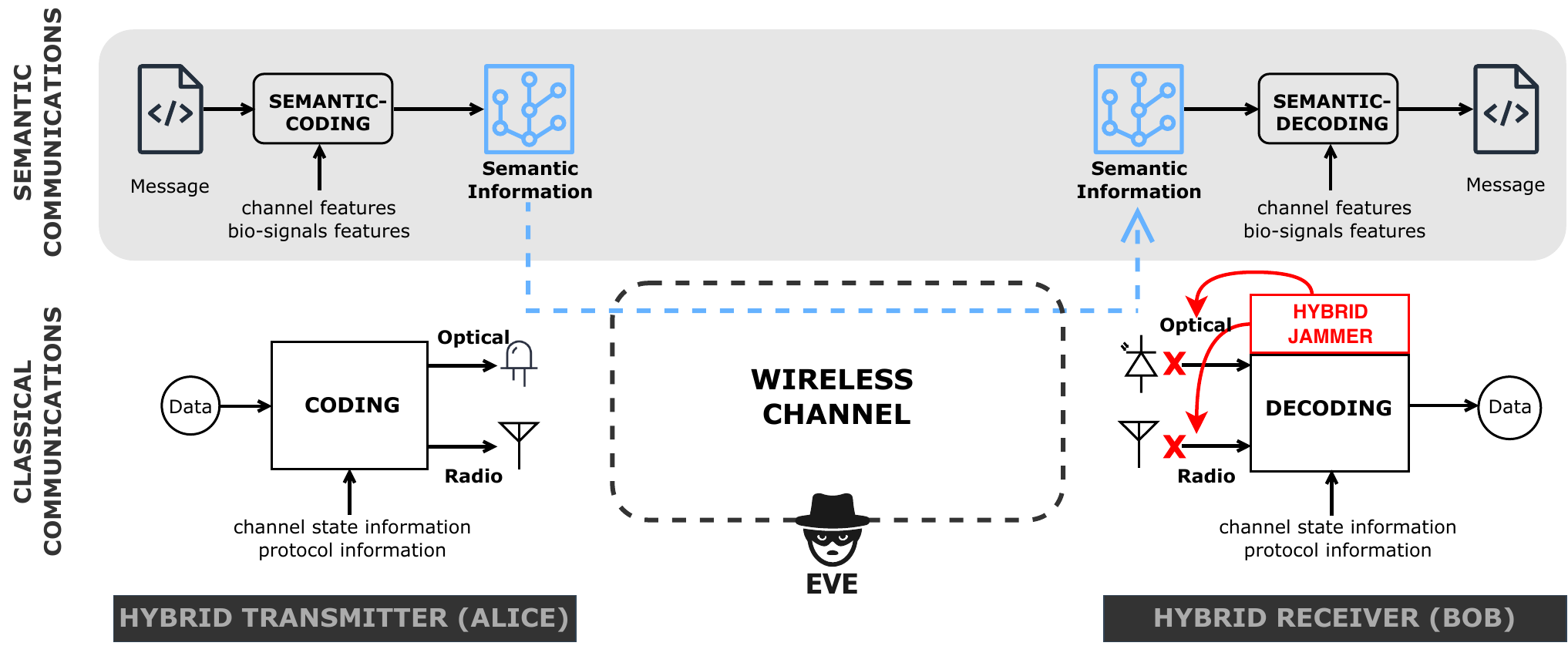}
	\caption{AI-based data stream encoding in \acp{hywban} using semantic communications and a jamming receiver to harden the security of wireless communications.}
	\label{fig:hywban-semantic}
\end{figure}

\simone{Understanding the implications for security in the dynamic landscape of \acp{hywban} is essential. Adversaries exploiting vulnerabilities in these networks could potentially gain unauthorized access to the human body, leading to critical threats like hijacking pacemakers, reconfiguring smart pill dispensers, or even creating novel types of diseases. The dual nature of these networks, encompassing radio and optical wireless channels, adds a layer of complexity to potential attacks. This study proposes a novel security mechanism combining the principles of semantic communications with the strategic deployment of a jamming receiver (see Fig.~\ref{fig:hywban-semantic}), enhancing the confidentiality and integrity of \acp{hywban}.}

\simone{Semantic communications~\cite{2021:beyond_shannon_emilio}, an emerging paradigm in network security~\cite{2023:semantic_iot}, involves generating semantic concepts related to biomedical applications or patient health status. This approach utilizes a \ac{dl} model trained on a dataset comprising measured, augmented, and synthetic biological signals~\cite{2023:smart-health_6G,2021:security_bionano_things}. During an enrollment phase, assumed free from adversarial presence, each semantic concept is associated with a secret, such as a cryptographic key, stored in the nodes' memory.}
%
%

\simone{The transmission of semantic concepts over the wireless channel, although susceptible to interception by malicious adversaries, is protected through a jamming receiver. As shown in Fig.~\ref{fig:hywban-semantic}, this receiver introduces intentional interference on either the light or radio channel, or both, effectively interfering with the transmitted data from the in-body device. Consequently, an adversary attempting to decode the data encounters altered signal characteristics, such as decreased \ac{snr} for radio and input power for \ac{nir} signals; this leads to an \textit{erroneous classification of semantic concepts}.}

\simone{However, the legitimate receiver, Bob, \textit{knows the jamming pattern} and can reverse the artificially induced bias to correctly decode the transmitted semantic concept. In contrast, an adversary, referred to as Eve, who lacks this knowledge, faces significant challenges in decoding the data accurately~\cite{2017:soderi_wbplsec,2022:soderi_pls_vlc_rgb}. This approach, leveraging semantic communication and controlled jamming, offers a dual-layered defence mechanism, enhancing the resilience of \acp{hywban} against sophisticated cyber threats.}
\simone{The rest of the section describes how the data were prepared and how we propose to use a \ac{dl} algorithm on devices with constrained resources.}

%
%
\subsection{Radio and optical channels data measurement} \label{subsec:data_meas}
\simone{In this study, we investigate the efficacy of hybrid communications that utilize optical and radio channels, explicitly investigating their capacity to penetrate biological tissues. Two distinct experimental setups were designed to assess the performance characteristics necessary for effective and secure communications through such mediums. For optical communications, \ac{nir} frequencies were employed, selected for their proven proficiency in penetrating biological tissues. On the other hand, \ac{uwb} technology, recognized for its superior transmission capabilities, particularly noise-like signals, was chosen for radio communications. This dual-faceted approach allows for a comprehensive evaluation of the potential of hybrid communication systems in medical applications.}

%
%
\begin{figure}[!ht]
	\centering
	\includegraphics[width=0.95\columnwidth]{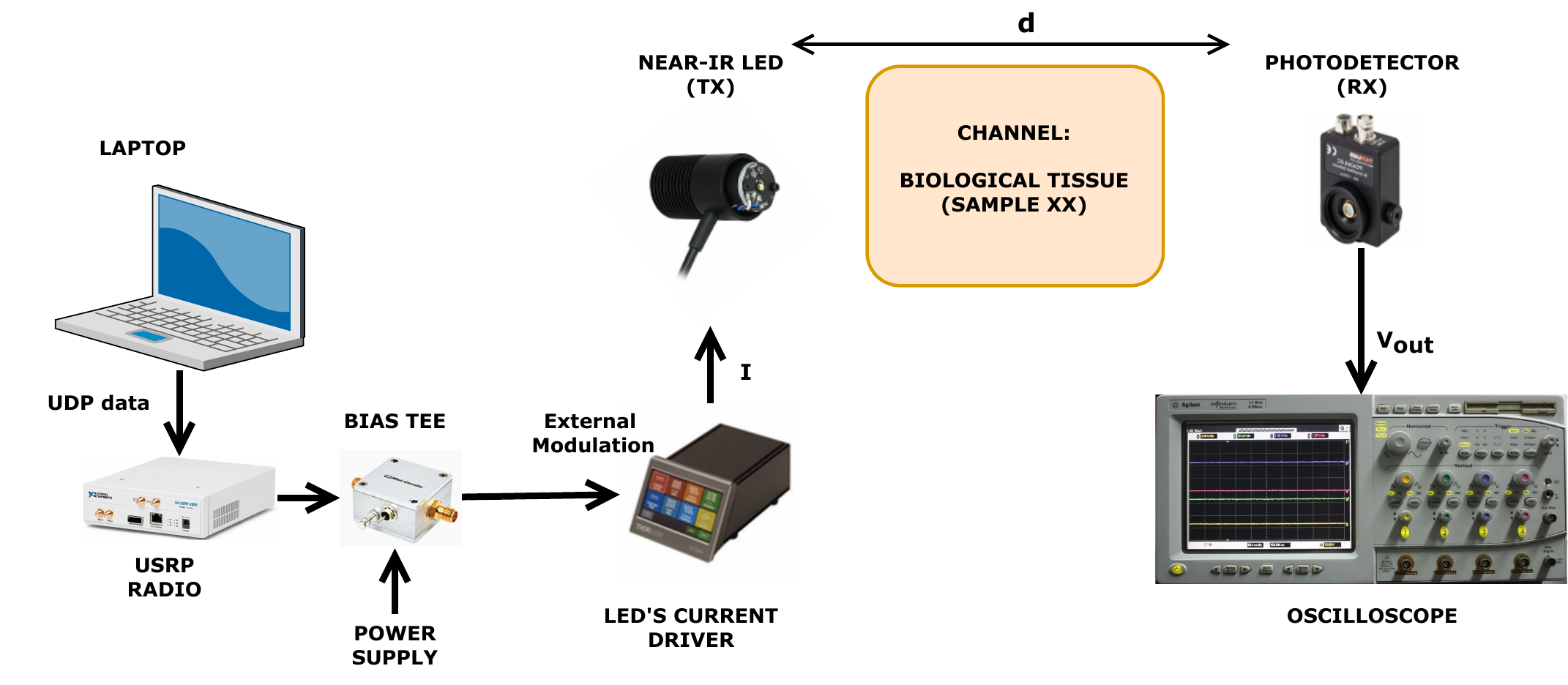}
	\caption{Measurement set-up with \ac{nir} communications (as \ac{owc}) through the biological tissue.}
	\label{fig:light-meas}
\end{figure}
\simone{The experimental setup, depicted in Fig.~\ref{fig:light-meas}, is an optical communication part; it comprises various components that can be divided into two subsystems: transmitter and receiver front end. The transmitter unit includes a \ac{nir} LED (M810L3, THORLABS) with 810 nm wavelength, a bias-tee, and a LED driver. The receiver unit utilizes a \ac{pd} (PDA 36A-EC  switchable gain detector, THORLABS). A sample of biological tissue was used, acting as the communications channel. The LED is driven by a current driver module (DC2200, THORLABS), which is controlled by an external modulation source. The modulation of the \ac{nir} LED is essential for transmitting data through the biological tissue. The \ac{pd}, positioned at a specific distance ($d$ is the thickness of the meat sample used in the measurements) from the \ac{nir} LED, captures the transmitted light after it has passed through the tissue. The output from the \ac{pd} ($V_{out}$) is then analyzed using an oscilloscope (with $50~\Upomega$ impedance) to assess the effectiveness of data transmission through the tissue sample. 
Using a laptop, we sent the same ASCII character with the \ac{udp} protocol to a software-defined radio USRP that modulated the signal before sending it to the LED driver. Two NI USRPs (2920 model) were employed in this study. We also used a bias-tee (ZFBT-4R2GW-FT+, Mini-Circuits) to combine the modulation signal and the bias current to feed the driver. We measured the peak of the received burst signal for each character sent from the laptop. The $V_{out}$ is then converted into an input power unit by using the equation provided in the datasheet of PDA36A-EC. This setup is crucial for evaluating the feasibility of \ac{nir} communications in scenarios where signals need to penetrate biological tissues, such as in implantable medical device applications.}

%
%
\begin{figure}[!ht]
	\centering
	\includegraphics[width=0.8\columnwidth]{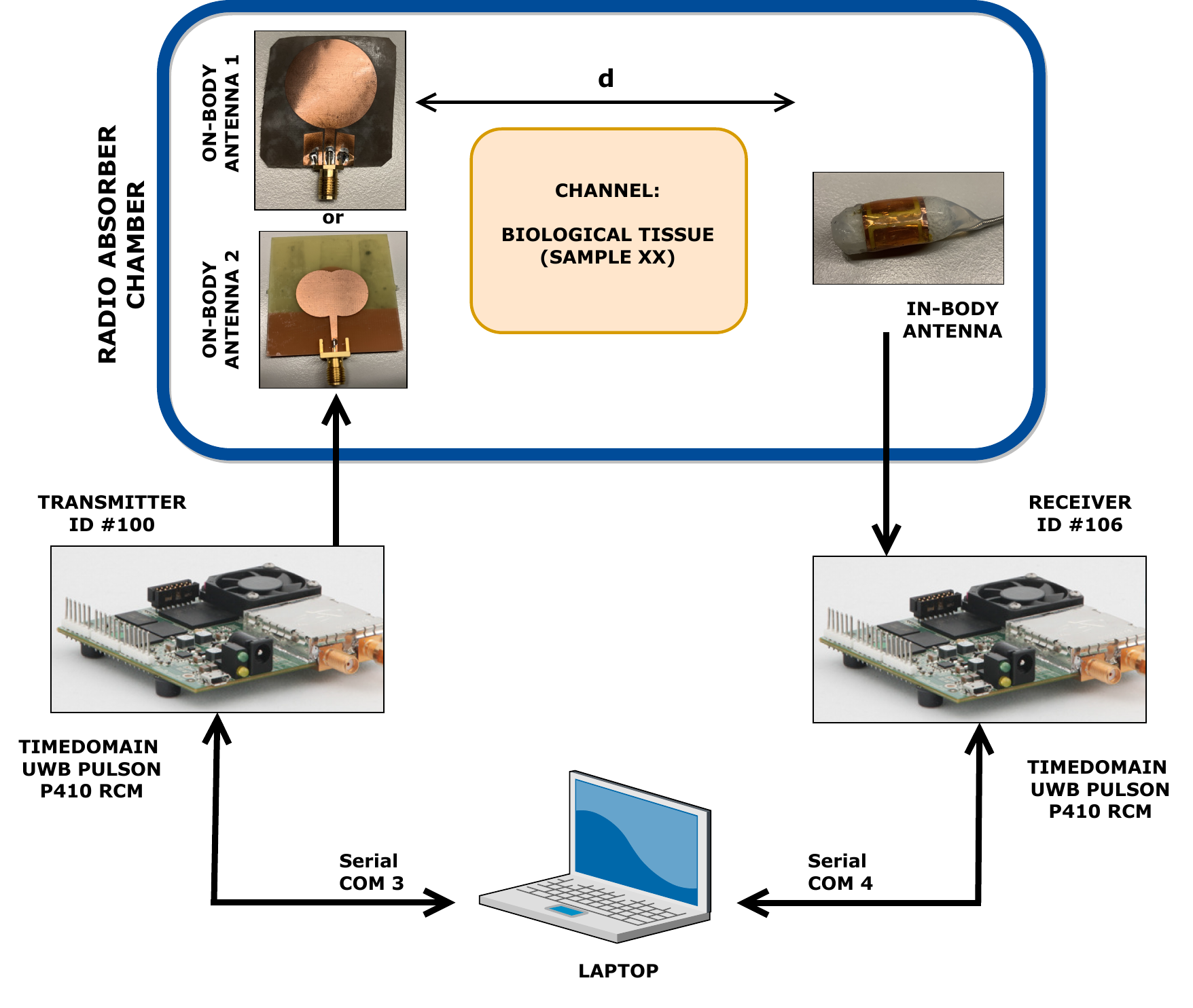}
	\caption{Measurement set-up with \ac{uwb} radio through the biological tissue.}
	\label{fig:radio-meas}
\end{figure}
\simone{As illustrated in Fig.~\ref{fig:radio-meas}, the radio measurement setup was meticulously designed to evaluate the performance of radio communications in \acp{hywban}. It consists of a \ac{uwb} transmitter (P410 PulsON, Time Domain) inside the body (i.e., in-body device) that communicates with a \ac{uwb} receiver (P410 PulsON, Time Domain) that has its antenna positioned on the porcine skin (i.e., on-body). Using Time Domain's Channel Analysis Tool (CAT) software, we simulated the communications scenario inside the body by sending signals from the transmitter to the receiver. We enclosed the antennas inside a box of RF absorber material to avoid external interference. The received signals were saved on a laptop using CAT software and analyzed later using MATLAB. This system makes it possible to accurately measure the radio signal's capability to penetrate biological tissue and the effectiveness of \ac{uwb} technology in an in-body communications scenario, essential for developing reliable hybrid WBANs.}

\begin{figure}[!t]
	\centering
	\includegraphics[width=0.75\columnwidth]{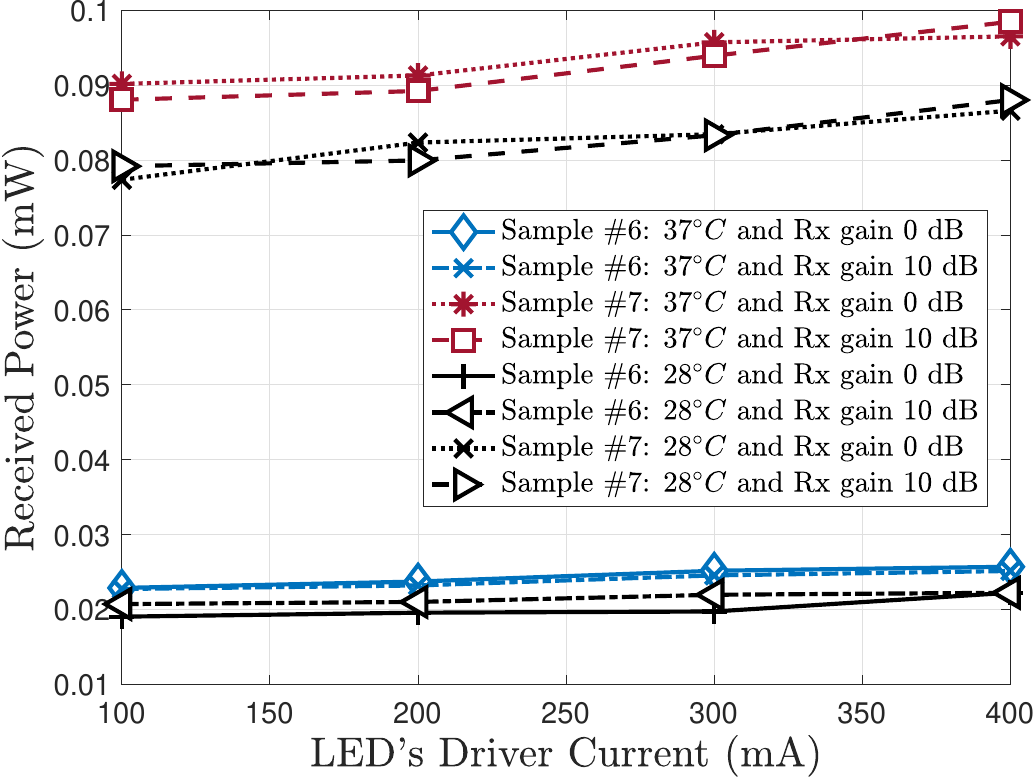}
	\caption{\ac{nir} received power varying the temperature of two biological tissue samples (i.e., sample \#6 and \#7) and the gain of the \ac{pd} .}
	\label{fig:nir-rx-power}
\end{figure}
\simone{Fig.~\ref{fig:nir-rx-power} shows the power (expressed in mW) of the \ac{nir} signal received after passing through the two biological tissue samples with a maximum thickness of $37$~mm and $39$~mm for samples \#6 and \#7, respectively.
From Fig.~\ref{fig:nir-rx-power}, it is evident how the propagation capabilities improve when the temperature reaches $37^\circ C$, which is considered almost typical for a human body. Selecting a higher gain (i.e., from $0$~dB to $10$~dB) in the receiver (PDA 36A-EC offers this option using a rotary switch) does not lead to a significant advantage regarding receiver sensitivity.}

%
%
\begin{figure}[!ht]
\centering
\subfigure[SNR with on-body antenna 1 and sample \#6 at $37^{\circ}$C.]
{
\includegraphics[width=0.75\columnwidth]{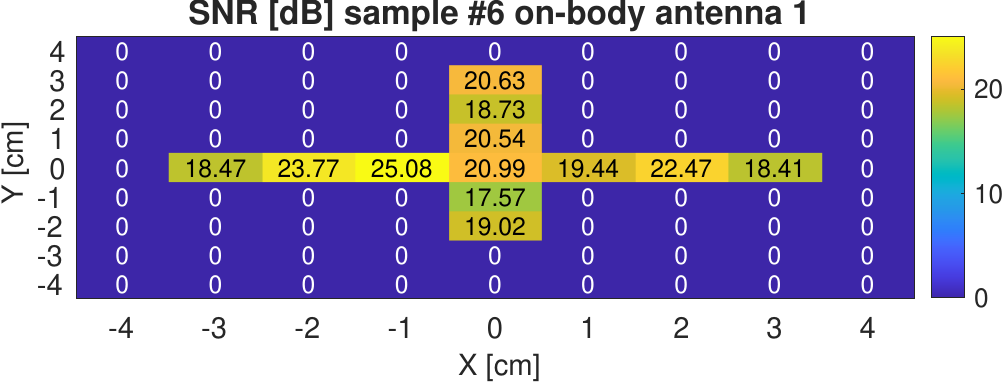}
\label{fig:uwb-sample6}
}
\\
\subfigure[SNR with on-body antenna 1 and sample \#7 at $37^{\circ}$C.]
{
\includegraphics[width=0.75\columnwidth]{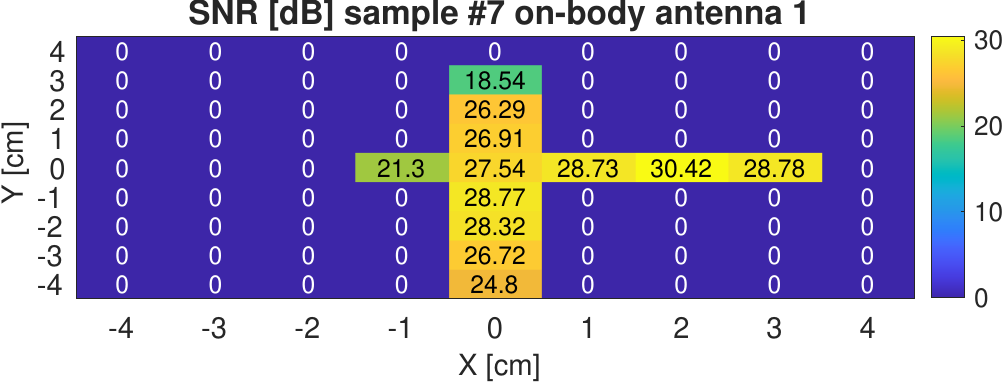}
\label{fig:uwb-sample7}
}
\caption{\ac{snr} measurements of \ac{uwb} transmissions through biological tissue.} 
\label{fig:uwb-samples} 
\end{figure}
\simone{Fig.~\ref{fig:uwb-samples} shows the \ac{uwb} \ac{snr} measured by the on-body antenna placed on the skin.  Meanwhile, the transmitter and receiver were aligned for light. For the \ac{uwb} measurements, we left the in-body antenna at the fixed position while we moved the on-body antenna in $1$~cm steps to investigate the communication limit. }

%
%
\subsection{Dataset: measured and synthetic data for medical applications using HyWBAN} \label{subsec:dataset}
\simone{Developing and optimizing semantic communications and strengthening security within hybrid networks necessitate a comprehensive dataset for training and testing \ac{dl} models. This dataset encompasses both measured features, such as \ac{snr} and the power received by the \ac{pd} in \ac{nir} communications, as well as synthetic features. These synthetic features are conceptualized on the premise that the constituent devices of \ac{hywban} can acquire and process biological signals from individuals. This dual approach in dataset formulation facilitates a realistic assessment of the \ac{hywban}'s operational capabilities and aids in simulating a wide range of scenarios for advanced medical applications.}

\simone{To refine the dataset for \ac{dl} models in the context of \acp{hywban} for medical applications, we employed a dual-strategy approach involving both the \textit{augmentation of measured data} and the \textit{generation of synthetic data}.
The augmentation process for the measured attributes, specifically \ac{snr} for \ac{uwb} and received power for \ac{nir}, employs a statistical methodology in which new values are generated based on a Gaussian distribution. This distribution is centred on the measured mean and standard deviation. This statistical rigour ensures the augmented data are closely aligned with realistic measurement variations.
We generated values for a few parameters: acceleration, heart rate and body temperature to generate synthetic data to emulate the ability of \ac{hywban} devices to measure biological signals. We assumed that a hybrid device could access these quantities (as a knowledge base for semantic communications) in a medical application or at least a part of it. These synthetic values are derived from a Gaussian distribution, adhering to predefined mean and standard deviations to ensure they fall within physiologically plausible ranges. 
The Table~\ref{tab:data-params} summarises the specification of statistics for the data generation process.}
%
%
\begin{table}[!b]
\centering
\small
\caption{Statistical summary of augmented and synthetic features.}
\label{tab:data-params}
\renewcommand{\arraystretch}{1.1}
\begin{threeparttable}
\begin{tabular}{cccc|c}
\hline
\textbf{Feature} & \textbf{Mean} & \textbf{Stand. Deviation} & \textbf{[Min, Max]} & \textbf{Threshold} \\
\hline
SNR (dB)\tnote{a} & 23.6 & 4.23 & [17.57, 33.32] & 19 dB\\
Input Power (mW)\tnote{b} & 0.07 & 0.03 & [0.02, 0.09] & 0.05 mW\\
\hline
Acceleration (m/s\textsuperscript{2}) & 0 & 0.1 & [-0.5, 0.5] &  0.1 m/s\textsuperscript{2}\\
Heart Rate (bpm) & 60 & 25 & [50, 120] & $<$60 bpm, $>$110 bpm\\
Body Temperature (°C) & 36 & 2 & [34, 42] & 37 °C\\
\hline
\end{tabular}
\begin{tablenotes}
\small
\item[a] \ac{uwb} measured data.
\item[b] \ac{nir} measured data.
\end{tablenotes}
\end{threeparttable}
\end{table}
\simone{Our semantic communication model uses a binary classification approach, simplifying complex data into categories like 'HIGH\_SNR' or 'LOW\_SNR' using the thresholds defined in Table~\ref{tab:data-params}. This method efficiently filters out noise, focusing on key data aspects and significantly reducing computational load. This binary representation accelerates model training and enhances interpretability, facilitating fast and decisive communications analysis.
We can then define the labels to be associated with each type of communication in a supervised manner (see Table~\ref{tab:communication_labels}). These labels are the \textit{semantic concepts} that represent data the device measures in a compressed manner.}
\simone{This particular approach to dataset preparation supports the robustness of the developed models. It ensures the simulation of diverse scenarios, which is critical for applying \ac{hywban} in medical settings.}

\begin{table}[!t]
\centering
\small
\caption{Classification labels for semantic communications}
\label{tab:communication_labels}
\begin{tabular}{cc}

\hline
\textbf{Label}            & \textbf{Condition}                              \\ \hline
\textbf{Full Communications        }& HIGH\_SNR and HIGH\_LPW                           \\ 
\textbf{Wide Communications}        & HIGH\_SNR and LOW\_LPW                            \\ 
\textbf{Communications in Motion }  & (HIGH\_SNR or HIGH\_LPW) and HIGH\_ACC            \\ 
\textbf{Critical Communications}    & (HIGH\_HR or HIGH\_TMP) and  LOW LPW            \\ 
\textbf{Unstable Communications}    & LOW\_SNR or LOW\_LPW                              \\ 
\textbf{Reduced Communications}     & Other scenarios not covered by above conditions   \\ 
\hline
\end{tabular}
\end{table}

%
%
\section{Proposed Model Evaluation} \label{sec:results}
Our study developed a deep learning model for semantic communication in \ac{hywban}: an autoencoder with a $64-32-64$ neuron structure and a classification model with dense layers and dropout regularization. 
The purpose of the autoencoder model within our semantic analysis is to reduce the dimensionality of the input data, including \ac{snr} and heart rate, thereby enabling more efficient processing and transmission. The autoencoder helps identify the most significant features crucial for semantic analysis by transforming the data into a lower-dimensional space. This process not only aids in preserving essential information but also contributes to the system's security by minimizing the amount of data exposed to potential threats.

We have conducted a series of experiments to determine the optimal architecture for our model. Our choice was guided by a grid search approach, where we evaluated various configurations and selected the one that minimized the reconstruction error on a validation set: the 64-32-64 structure balanced model complexity and the ability to capture the underlying patterns in the data. We also experimented with different activation functions and learning rates, ultimately choosing a Rectified Linear Unit (ReLU) activation for its efficiency and a learning rate of 0.001 for stable convergence.
The training process of our autoencoder was carried out over $50$ epochs, with early stopping implemented to prevent overfitting. We used a batch size of $256$, which was determined to be optimal through experimentation, balancing the trade-off between training speed and memory constraints. The dataset (after an augmentation of the data by a factor of $50$) comprised $2040$ samples, split into $80$\% for training and $20$\% for validation. This information aims to enhance the transparency and reproducibility of our model evaluation.

\begin{figure}[!h]
	\centering
	\includegraphics[width=0.8\columnwidth]{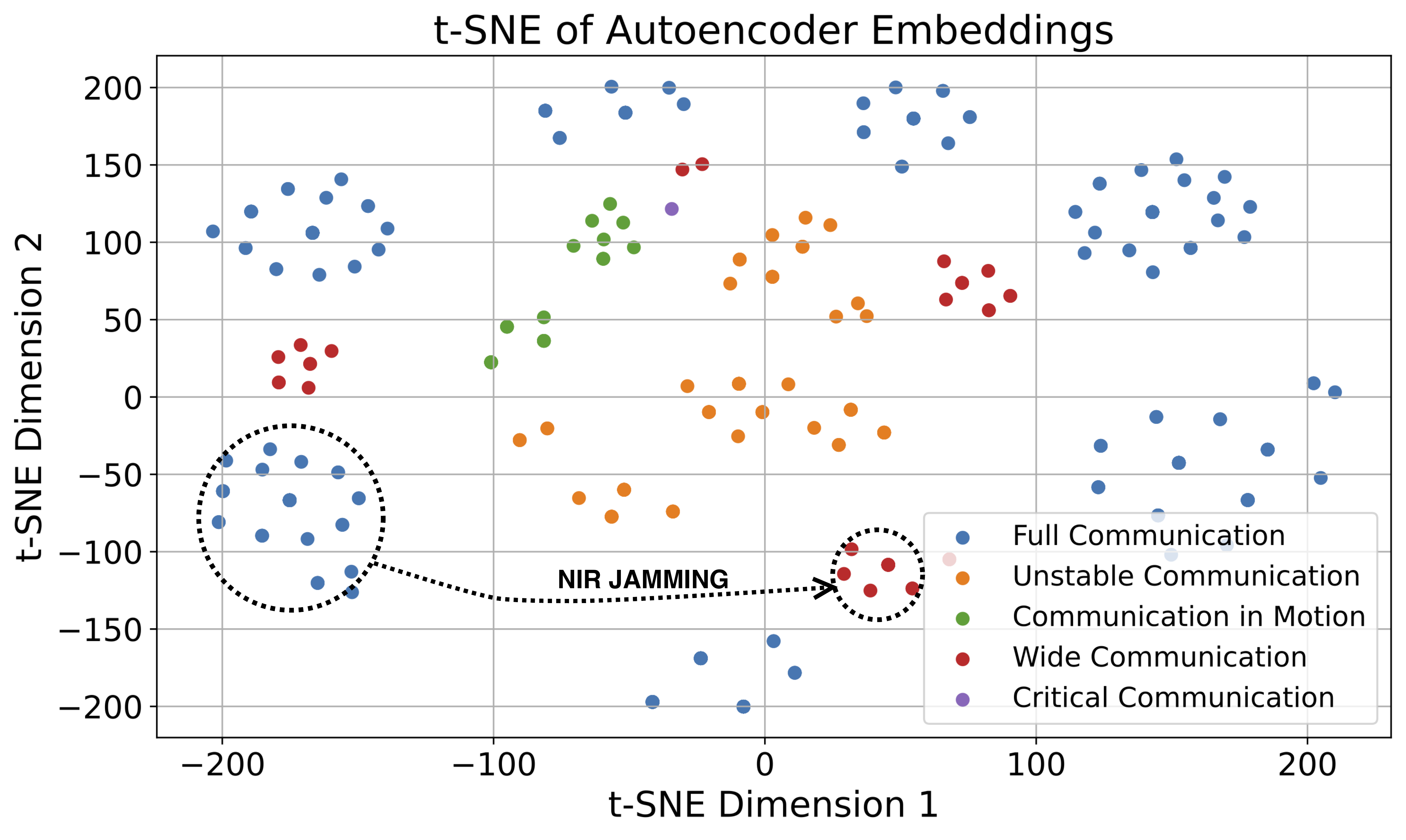}
	\caption{Low-dimensional representation of data preserving semantic similarities. For example, the figure shows the effect of \ac{nir} jamming on the semantic concepts.}
	\label{fig:tSNE-view}
\end{figure}

To visualize the effectiveness of the autoencoder in capturing semantic relationships, we apply the t-Distributed Stochastic Neighbor Embedding (t-SNE) technique (see Fig.~\ref{fig:tSNE-view}). This method is noted for its ability to represent high-dimensional data in lower dimensions while preserving data structures, allowing us to visually inspect the clustering of data points based on their semantic similarities.
We have expanded our discussion on the interpretability of clusters formed in the t-SNE visualization. The clusters represent distinct data patterns that the autoencoder has learned to encode. By examining the characteristics of samples within each cluster, we can infer the model's ability to discern different features in the data, which supports its effectiveness.

The model is optimized using \textit{Adam} optimizer and trained to categorize the data into predefined semantic classes, as described in our data preprocessing phase. Performance evaluation using a confusion matrix and accuracy metrics confirmed the model's efficacy. Finally, the model was converted to \textit{TensorFlow Lite}~\cite{2021:tinyml-iot-survey} (i.e., a Tiny Machine Learning framework that supports the conversion of ML models into a format that can be run on microcontrollers), aligning with low-power, edge-based IoT device requirements, ensuring privacy, energy efficiency, and real-time processing. This approach signifies a substantial advancement in semantic communication for smart healthcare applications.

%
%

\begin{figure}[!t]
	\centering
	\includegraphics[width=0.8\columnwidth]{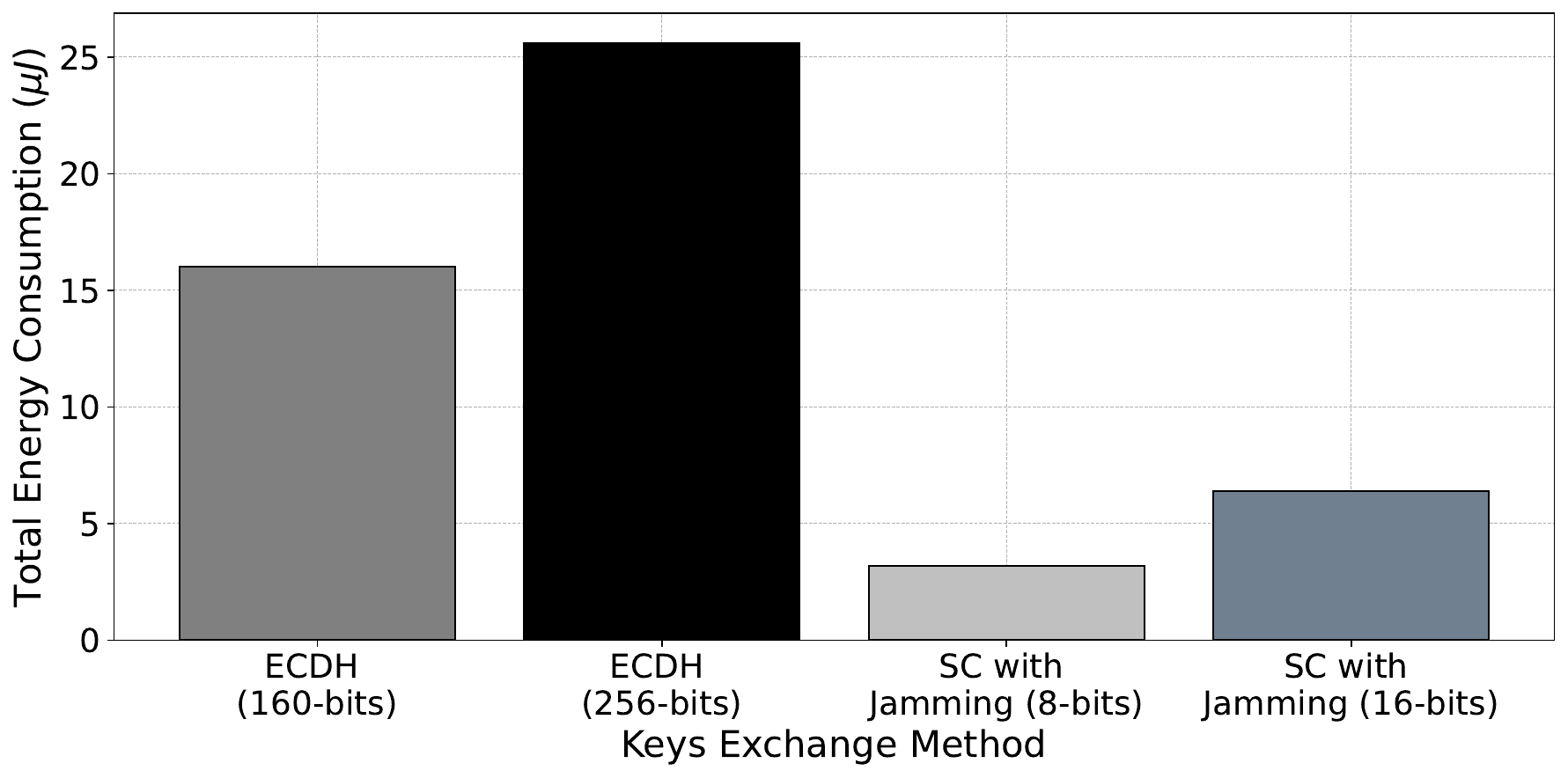}
	\caption{Energy consumption comparison between \ac{sc} with jamming and ECDH.}
	\label{fig:energy-sc}
\end{figure}

\simone{In our evaluation, we compared the energy efficiency of our semantic communication with a jamming solution to the Elliptic Curve Diffie-Hellman (ECDH) key exchange~\cite{2017:crypto-net-sec-Stallings-7th}. The comparison focused on various configurations, assessing the energy consumption for different key lengths in ECDH ($160$ and $256$ bits) against our semantic communication model that can use $8$ or $16$ bits to represent the semantic concepts, and it is enhanced with jamming up to $8$ and $16$ bits (i.e., worst case for our proposal). We assumed $0.1~\mu$J as the energy per bit and $0.2~\mu$J as the energy cost to jam a bit. This analysis, crucial for understanding the practicality of deploying these methods in energy-constrained environments like \acp{hywban}, is visualized Fig.~\ref{fig:energy-sc}, illustrating the total energy consumption of each method. Such comparisons highlight the efficiency of semantic communications, especially when supplemented with jamming, in contrast to traditional cryptographic approaches like ECDH.}

\simone{The classification performance of our TensorFlow model for semantic communication in \acp{hywban} demonstrated good precision and recall across most classes, with an overall accuracy of $94$\%. However, the corresponding TensorFlow Lite model, optimized for low-power devices, showed a variation in performance, particularly in precision and recall for specific communication classes like \textit{Communication in Motion} and \textit{Critical Communication}, resulting in an overall accuracy of $86$\%. This variation underscores the challenges in collecting more data with measurements and balancing model complexity with the constraints of edge computing devices.}
%
%

\section{Conclusions}  \label{SEC:Conclusions}
The research presented in this paper marks a significant stride in enhancing the security of \acp{hywban}. By integrating semantic communications with jamming receivers, we demonstrate a robust method to protect sensitive health data and biomedical devices within the \ac{hywban} framework. Our experimental analysis provides valuable insights into the propagation characteristics of hybrid communications in biological tissues, forming the basis for an advanced \ac{dl} model. This model's ability to generate and interpret semantic concepts, coupled with a strategic jamming mechanism, ensures the reliable transmission of encrypted data, thereby mitigating potential cybersecurity threats. Notably, our approach outperforms traditional cryptographic methods in energy efficiency, making it a viable solution for the energy-sensitive environment of HyWBANs. 

The semantic strategy enhances security by transmitting only the necessary and relevant data, reducing the attack surface. The deep learning model contributes to this by learning to identify and filter out non-essential information, thus streamlining the communication process and making it more secure. The inherent security advantages of optical communications, such as the line-of-sight requirement for the interception, are exploited in our hybrid system to strengthen overall security further.

The findings and methodologies outlined in this study improve the security of current \ac{hywban} systems and pave the way for their broader adoption in smart healthcare services, aligning with the evolving landscape of 6G technology.

%
%
\section{Acknowledgments}\label{SEC:Acks}

This research is partially funded by the Research Council of Finland Profi6 funded 6G Enabling Sustainable Society (University of Oulu) and Research Council of Finland 6G Flagship (grant 318927).
and the European Union’s Horizon 2020 programme under the Marie Sklodowska-Curie grant H2020-MSCA-RISE-2019 agreement No. 872752.
%
%
%
\bibliographystyle{splncs03_unsrt.bst} 
\bibliography{bibliography}

\end{document}